\documentclass[10pt, conference, compsocconf]{IEEEtran}
\IEEEoverridecommandlockouts
\usepackage{bbding}
\usepackage{url}
\usepackage{booktabs}

\usepackage{CJK}
\usepackage{subfigure}
\usepackage{amsmath,amssymb,amsthm}
\usepackage{multirow}
\usepackage{fancybox}
\usepackage[dvipsnames,svgnames,x11names]{xcolor}
\usepackage{graphicx}
\usepackage{bbm}
\usepackage{algorithm} 
\usepackage{algorithmic} 
\usepackage{verbatim}
\usepackage{booktabs}
\usepackage[UKenglish]{babel}
\usepackage{setspace}
\makeatletter
\newcommand*{\rom}[1]{\expandafter\@slowromancap\romannumeral #1@}
\makeatother
\newcommand\norm[1]{\left\lVert#1\right\rVert}

\DeclareMathOperator*{\argmax}{argmax}

\ifCLASSINFOpdf
\else
\fi
\begin{document}

%
\title{A Consumer Behavior Based Approach to Multi-Stage EV Charging Station Placement}

\author{\IEEEauthorblockN{Chao Luo,
Yih-Fang Huang, and Vijay Gupta}
\IEEEauthorblockA{Department of Electrical Engineering\\
University of Notre Dame, Notre Dame, Indiana, USA\\
Email: {\{cluo1, huang, vgupta2\}@nd.edu}}
\thanks{This work has been partially supported by the National Science Foundation under grants CNS-1239224 and ECCS-0846631.}
}



%


\maketitle

\begin{abstract}
This paper presents a multi-stage approach to the placement of charging stations under the scenarios of different electric vehicle (EV) penetration rates. The EV charging market is modeled as the oligopoly. A consumer behavior based approach is applied to forecast the charging demand of the charging stations using a nested logit model.  The impacts of both the urban road network and the power grid network on charging station planning are also considered. At each planning stage, the optimal station placement strategy is derived through solving a Bayesian game among the service providers. To investigate the interplay of the travel pattern, the consumer behavior, urban road network, power grid network, and the charging station placement, a simulation platform (The EV Virtual City 1.0) is developed using Java on Repast. We conduct a case study in the San Pedro District of Los Angeles by importing the geographic and demographic data of that region into the platform.  The simulation results demonstrate a strong consistency between the charging station placement and the traffic flow of EVs.  The results also reveal an interesting phenomenon that service providers prefer clustering instead of spatial separation in this oligopoly market.
\end{abstract}

%
\IEEEpeerreviewmaketitle

\section{Introduction}
\begin{spacing}{0.92}
Recent advances in EV technology are leading us into an era of large-scale vehicle electrification. However, the sparsity of current public charging infrastructure remains as a major impediment to the proliferation of EVs. Our work aims to develop guidelines for the charging service providers to make a foresightful decision on charging station placement so they have a good chance to make profits.

In our work, multiple charging service providers try to maximize their own overall utility while satisfying the Quality-of-Service (QoS) constraints when choosing the optimal charging station placement. Our approach is based on a multi-stage planning strategy, where the service providers need to forecast the charging demand at a charging station candidate at each planning stage. Characterizing the charging demand also requires consumer behavior analysis to reflect the various consumer preferences. To this end, we employ the nested logit model to analyse the charging behavior of EV owners. Then, we derive the optimal placement strategy by solving a Bayesian game among the service providers. Finally, we develop a simulation platform using Java on Repast \cite{Repast}, and conduct a case study in the San Pedro District of Los Angeles.

There is an increasing literature aimed at addressing the issue of EV charging infrastructure deployment. \cite{Shaoyun}-\cite{Albert} have formulated charging station placement as an optimization problem. However, the objective functions in those optimization problems did not consider the consumer's overall satisfaction (utility) in terms of charging price, cost of travel to the station, and amenities available near the charging station (restaurant, supermarket, etc.) Furthermore, the optimization framework was constructed from the perspective of a central planner instead of providers in a deregulated market. Bernardo \emph{et al.} \cite{cite9} employed a discrete choice model (DCM) to investigate the optimal locations for fast charging stations. They modeled each potential EV charging station as a player in the game. However, their formulation assumed that each player has the complete information about the other players. This assumption may be too strict and implausible in a realistic market. In our work, the Bayesian game is posed with an information structure that supposes incomplete information among players.

The main contributions of our work are as follows: (1) A multi-stage charging station deployment framework with different EV penetration rates is presented; (2) A nested logit model is employed to analyze the consumer's satisfaction and forecast the charging demand, which provides us insights into the behavioral process of EV owners' charging decisions; (3) An oligopolistic market model of EV charging service providers and EV owners is studied using a Bayesian game among multiple charing service providers; (4) A Java simulation platform (The EV Virtual City 1.0) has been developed to analyze the interplay of travel patterns of EV owners, charging demand, urban infrastructure and charging station placement strategies.

\section{Problem Formulation}
\begin{table*}[htbp]
\center
\caption{Charge Method Electrical Ratings}\label{ta1}
 \begin{tabular}{cccc}
  \toprule
  Charging Method& Nominal Supply Voltage(Volts)&Maximum Current (Amps) & Time from fully depleted to
fully charged \\
  \midrule
 Level 1 &120 vac, 1-phase &12 A & 16-18 hours \\
 Level 2 & 208 to 240 vac, 1-phase&32 A & 3-8 hours \\
 Level 3&600 vdc maximum&400 A maximum & less than 30 minutes  \\
  \bottomrule
 \end{tabular}
\end{table*}
We consider the special case of three service providers which offer three EV charging services \cite{specification}, namely, Level 1, Level 2, and Level 3 (see Table \ref{ta1}). This setting can be easily generalized to more service providers. Specifically, we assume that service provider 1 offers Level 1 charging service, service provider 2 offers Level 2 charging service, and service provider 3 offers Level 3 charging service. Each service provider can, however, own multiple charging stations. At each planning stage, the three service providers choose some locations from a given set of candidate locations, denoted as $\mathcal{I}=\{1,2,3\cdots,L\}$,  to place the charging stations. Let $\mathcal{E}=\{1,2,3,\cdots,N\}$ index the EVs.

\subsection{The Profit of Charging Service Provider}
We assume that charging stations affiliated with the same service provider have the same retail price. Each charging station purchases the electricity from the wholesale market at the locational marginal price (LMP). In a deregulated electricity markets, LMP is calculated for every node by market coordinator \cite{PJM}. Let $p_k$ represent the retail price of provider $k\;(k=1,2,3)$, and $c_{j,k}$ represent the LMP of the $j$th charging station candidate of service provider $k$. Let $p_{-k}$ denote the retail prices of the other two service providers except $k$, and $\psi_{j,k}$ denote the expected charging demand of the $j$th charging station candidate of service provider $k$. Define $S_k=[s_{1,k},s_{2,k},\cdots, s_{L,k}]^{\textrm{T}}$ as the placement decision of service provider $k$, where $s_{j,k}\in\{0,1\}$ is a binary variable with $s_{j,k}=1$ indicating service provider $k$ will place a charging station at the $j$th candidate location, and $s_{j,k}=0$ otherwise. Let $S_{-k}$ represent the placement decisions of the other two service providers except $k$. Define $\theta_{j,k}$ as the setup cost of the $j$th charging station candidate. Define $\Pi_k$ as the total profit of service provider $k$.

\begin{equation}
\Pi_k=p_k\Psi_k^{\textrm{T}}S_k-\textrm{diag}[C_k]\Psi_k^{\textrm{T}}S_k
-\Theta_k^{\textrm{T}}S_k,
\end{equation}
and define
\begin{equation}
R_k=p_k\Psi_k^{\textrm{T}}S_k-\textrm{diag}[C_k]\Psi_k^{\textrm{T}}S_k,
\end{equation}
where $\Psi_k=[\psi_{1,k},\psi_{2,k},\psi_{3,k},\cdots,\psi_{L,k}]^{\textrm{T}}$, and $C_k=[c_{1,k},c_{2,k},\cdots,c_{L,k}]^{\textrm{T}}$ and $\Theta_k=[\theta_{1,k},\theta_{2,k},\cdots,\theta_{L,k}]^{\textrm{T}}$. $\textrm{diag}[.]$ is an operator to create a diagonal matrix using the underlying vector. $[.]^{\textrm{T}}$ is the transpose operation. $p_k\Psi_k^{\textrm{T}}S_k$ is the total sales,  $\textrm{diag}[C_k]\Psi_k^{\textrm{T}}S_k$ is the cost of purchasing electricity, and $\Theta_k^{\textrm{T}}S_k$ accounts for the setup cost. $R_k$ is the total revenue of service provider $k$.

\subsection{The Impact of EV Charging on Power Grid}
Large-scale EV integration will present many challenges on the power grid, e.g. system stability, power loss, frequency regulation, etc \cite{impact1}-\cite{impact3}. For the power grid, the generators will collaboratively adjust the output of real power and reactive power to maintain system stability, perform frequency regulation or voltage regulation. The 2-norm deviation of generating power (real power and reactive power) is widely used as a metric to evaluate the difficulty in mitigating the ``disturbance" of the power grid caused by external factors. In our case, the 2-norm difference between the generating power with and without EV charging is used as a metric to characterize the impacts of EV charging on the electric power system. Assuming that the power system has $M$ generators and $D$ buses (substations), we define the disturbance as

\begin{equation}
B=\norm{P_g^{\textrm{base}}-P_g^{\textrm{EV}}}^2_2+
\norm{Q_g^{\textrm{base}}-Q_g^{\textrm{EV}}}^2_2,
\end{equation}
where $P_g^{\textrm{base}}$ is a $M\times1$ vector representing the real power generated by the $M$ generators under base power load scenario (without EV charging), and $P_g^{\textrm{EV}}$ is the vector of real power under the EV charging scenario (with EV charging).  Similarly, $Q_g^{\textrm{EV}}$ and $Q_g^{\textrm{base}}$ are the vectors of reactive power with and without EV charging, respectively.

\subsection{Quality-of-Service Constraints}
We propose two quality-of-service (QoS) metrics: (1) average service delay probability $\Upsilon_k,\;(k=1,2,3)$, (2) average service coverage $\Xi_k,\;(k=1,2,3)$.

\begin{equation}
\Upsilon_k=\frac{1}{N}\sum_{i=1}^N\upsilon_{i,k}\;\;\;\;\;\;\Xi_k=\frac{1}{N}\sum_{i=1}^N\xi_{i,k},
\end{equation}
where $\upsilon_{i,k}$ is the average service delay probability of the $i$th EV owner with respect to service provider $k$. $\xi_{i,k}$ is the average number of accessible Level $k$ charging stations when the $i$th EV owner travels around.

\subsection{Multi-stage Charging Station Planning Scheme}
We define the utility function of service provider as follows:
\begin{equation}
U_k=\Pi_k-wB_k,
\end{equation}
where $w$ is a weight coefficient. The first term is the total profits made from EV charging, and the second term characterizes the penalty of power grid from large-scale EV charging.

At each new stage, the service providers obtain the optimal placement strategy by solving the following problem.

\begin{equation}\label{obj}
\begin{aligned}
&[S_k^T|S_1^{T-1},S_2^{T-1},S_3^{T-1}]=\\
&\argmax_{\substack{{s_{1,k},\cdots,s_{L,k}}\\{s_{j,k}\in\{0,1\}}}}\left\{\mathbb{E}_{S_{-k}}[U_k]|S_1^{T-1},
S_2^{T-1},S_3^{T-1}\right\},
\end{aligned}
\end{equation}
subject to
\begin{equation}
\Upsilon_k\leq\Upsilon^0,\;\;\;\;\;\Xi_k\leq\Xi^0,
\end{equation}
where $\mathbb{E}_{S_{-k}}[.]$ is the expectation over $S_{-k}$. $S_k^T$ accounts for the placement decision at stage $T$.

To solve this problem, we need to answer two principal questions: (1) How to forecast the charging demand at each charging station candidate? (2) How to calculate the optimal placement strategy in a more effective way? In Section \rom{3}, we apply the nested logit model to estimate the charging demand.  In Section \rom{4}, we employ a Bayesian game model to characterize the strategic interaction among the service providers and derive the placement strategies.

\section{Charging Demand of EV Charging Station}

The charging demand at a charging station candidate is defined as the sum of the product of the probability that EV owners will go to that charging station and the energy needed to charge the EVs. Many factors may influence the charging behaviors, such as the retail charging price, travel distances, amenities available near the charging station, the travel purpose of EV owners, etc. In our work, we apply the nested logit model to characterize the attractiveness of a charging station and analyze the charging behavior of EV owners \cite{cite11}.

\subsection{Nested Logit Model And Probability of Choice}
The nested logit model belongs to the family of discrete choice model (DCM), which is widely utilized in the analysis and forecast of a consumer's decision among a finite set of choice alternatives \cite{cite11}. The main idea of DCM is that a consumer tries to maximize the total utility when making a decision on choosing from multiple choice alternatives.

In our problem, there are three service providers offering Level 1 charging, Level 2 charging and Level 3 charging. Each provider operates multiple charging stations. The utility that the $n$th EV owner can obtain from choosing charging station $j\;(j=1,2,\cdots,L)$ of service provider $k\;(k=1,2,3)$ is denoted as $U_{j,k}^n=\overline{U}_{j,k}^n+\epsilon_{j,k}^n$, where $\overline{U}_{j,k}^n$ corresponds to the observable utility and $\epsilon_{j,k}^n$ corresponds to the unobservable utility. The vector of unobservable utility $\epsilon^n=[\epsilon_{1,1}^n,\cdots,\epsilon_{L,1}^n,\epsilon_{1,2}^n,\cdots,\epsilon_{L,2}^n, \epsilon_{1,3}^n,\cdots,\epsilon_{L,3}^n]^{\textrm{T}}$ has a generalized extreme value (GEV) distribution with cumulative distribution function given by

\begin{equation}\label{cdf}
F(\epsilon^n)=\exp\left(-\sum_{k=1}^3\left(\sum_{l=1}^Le^{-\epsilon_{l,k}^n/\sigma_k}\right)^{\sigma_k}\right),
\end{equation}
where $\sigma_k$ is a measure of the degree of independence in the unobservable utility among the charging stations owned by service provider $k$.

For nested logit model, we can decompose the observable utility $\overline{U}_{j,k}^n$ into two components: the utility of choosing a service provider (i.e. charging level) and the utility of choosing a charging station. In addition, we assume home charging acts as the ``outside good" in the market \cite{outside1}-\cite{outside2}. Hence, $\overline{U}_{j,k}^n$ for EV owner $n$ can be expressed as

\begin{equation}
\overline{U}_{j,k}^n=\overline{W}_{k}^n+\overline{V}_{j,k}^n,
\end{equation}
where $\overline{W}_{k}^n$ corresponds to the observable utility of choosing service provider $k$ (choosing nest $k$), and $\overline{V}_{j,k}^n$ accounts for the observable utility of choosing charging station $j$ given that service provider $k$ has been chosen. $\overline{W}_{k}^n$ and $\overline{V}_{j,k}$ are linear weighted combinations of characteristics of both the charging stations and the EV owner.

\begin{equation}
\overline {W}_k^n =\gamma_{k,1}y_1^n+\gamma_{k,2}y_2^n,
\end{equation}
where $y_1^n$ and $y_2^n$ represent, respectively, the travel purpose and income of EV owner $n$, and $ \gamma_{k,1}, \gamma_{k,2}$ are the corresponding weight coefficients. As an ``outside good" in the market, the utility of home charging is normalized, i.e. $\overline{W}_0^n=0$.

\begin{equation}
\begin{aligned}
\overline{V}_{j,k}^n=&
\alpha_kp_k+\mu_kd_{j,k}^n+\eta_kz_{j,k}^n+\\
&\lambda_{k,1}x_{j,k,1}+\lambda_{k,2}x_{j,k,2}+\lambda_{k,3}x_{j,k,3},\\
\end{aligned}
\end{equation}
where $p_k$ is the retail charging price and $z_{j,k}^n$ is the destination indicator. If the $j$th charging station  is near the EV owner's travel destination (within a threshold distance $d_{th}$), $z_{j,k}^n=1$, otherwise, $z_{j,k}^n=0$. $d_{j,k}^n$ is the deviating distance due to EV charging. Additionally, each candidate charging station is associated with a set of characteristics $X_{j,k}=[x_{j,k,1}, x_{j,k,2}, x_{j,k,3}]^{\mathrm{T}}$, which characterizes the attractiveness of this charging station in terms of three amenities. For instance, if there exists a restaurant near location $j$, we set $x_{j,k,1} = 1$, otherwise $x_{j,k,1} = 0$. Similarly, $x_{j,k,2}$ and $x_{j,k,3}$ are the indicators for shopping center and supermarket, respectively. $\alpha_k,\mu_k,\eta_k,\lambda_{k,1},\lambda_{k,2},\lambda_{k,2}$ are weighting coefficients.

The EV owner's choice is not deterministic due to the random unobservable utility. However, we can derive the probability that he/she will choose a certain charging station by taking the expectation over the unobservable utilities defined in Equation (\ref{cdf}). The probability that the $n$th EV owner will choose the $j$th charging station of service provider $k$ is \cite{cite11}

\begin{equation}
\begin{aligned}
\Phi_{j,k}^n&=\mathbf{Prob}\left(\overline{U}_{j,k}^n+\epsilon_{j,k}^n > \overline{U}_{i,l}^n + \epsilon_{i,l}^n, \forall i \neq j, \textrm{or } l \neq k \right)\\
&=\int_{-\infty}^{+\infty}F_{j,k}(\overline{U}_{j,k}^n-\overline{U}_{1,1}^n+\epsilon_{j,k}^n,\overline{U}_{j,k}^n-\overline{U}_{2,1}^n+\epsilon_{j,k}^n,\\
&\cdots,\epsilon_{j,k}^n,\cdots,\overline{U}_{j,k}^n-\overline{U}_{L,3}^n+\epsilon_{j,k}^n)d\epsilon_{j,k}^n
\end{aligned}
\end{equation}
where $F_{j,k}$ denotes the derivative of $F$ with respect to $\epsilon_{j,k}^n$, i.e. $F_{j,k}=\partial F/\partial \epsilon_{j,k}^n$. Finally, we obtain

\begin{equation}
\Phi_{j,k}^n=\frac{e^{\overline{U}_{j,k}^n/\sigma_k}\left(\sum_{l=1}^Le^{\overline{U}_{l,k}^n/\sigma_k}\right)^{\sigma_k-1}}
{\sum_{t=1}^3\left(\sum_{l=1}^Le^{\overline{U}_{l,t}^n/\sigma_t}\right)^{\sigma_t}}.
\end{equation}

Once the EV owners' choice probability is calculated, one can estimate the charging demand of a charging station.  Let $q_n\;(n=1,2,\cdots,N)$ denote the total energy (measured in kWh) EV owner $n$ plans to purchase to charge the vehicle, and assume that $q_n$ is a random variable uniformly distributed in the range $[0.5Q, Q]$, where $Q$ denotes the battery capacity of EVs. To simplify the analysis, further assume that all EVs have the same battery capacity. The total charging demand for charging station $j$ of service provider $k$ is given by
\begin{equation}
\psi_{j,k}=\sum_{n=1}^Nq_n\Phi_{j,k}^n.
\end{equation}

All the weight parameters in the nested logit model can be estimated using the data from stated and revealed preference survey. The key component of nested logit model is to calculate the probabilities that an EV owner will go to the given charging stations. However,  it does not necessarily imply that each time an EV owner will always decide on which charging station according to the probabilities. An individual EV owner may still go to a fixed charging station regularly. The charging demand calculation is statistically meaningful only when we sum up the individual charging demand over a substantial number of EVs.

\section{Optimal Placement in A Bayesian Game}
In practice, a service provider usually does not know the exact setup costs and payoff functions of the other two providers, so we pose the problem as a Bayesian game \cite{bayesian} among the service providers at each stage of charging station planning. In this game, a player corresponds to a service provider.

For simplicity, we drop the stage index in the following definitions since the Bayesian game has the same scheme at each stage. A Bayesian game consists of a set of players $\mathcal{I}$, a strategy space $S_k$, a type space $\Theta_k$, a payoff function $u_k$ and the joint probability of the types $f(\Theta_1,\Theta_2,\Theta_3)$. $S_k=[s_{1,k},s_{2,k},\cdots, s_{L,k}]^{\textrm{T}}$ corresponds to all possible placement strategies for player $k$. $S_{-k}$ corresponds to the placement strategies of the other players except player $k$. We define $f(S_{-k})$ as probability mass function (PMF) of the placement strategies of the other players. The type space $\Theta_k=[\theta_{1,k},\theta_{2,k},\cdots,\theta_{L,k}]^{\textrm{T}}$ represents the setup costs of all charging station candidates of service provider $k$. In this paper, we assume that a service provider knows its own type, but not the exact types of the other two service providers.

Denote $\theta_{j,k}$ as the setup cost for charging station $j$ of service provider $k$, which includes the equipment cost, installation fee, construction cost, land rental, etc. $\theta_{j,k}$ has i.i.d. uniform distribution.

Before proceeding to solve the Bayesian game, we make the following assumptions.

$\mathbf{Assumption\;1}$: $f(S_{-k})$ is binomially distributed with parameter 0.5, \emph{i.e.} $S_{-k}\backsim \textrm{Binomial}(2L, 0.5)$.

 $\mathbf{Remark\;1}$: The distribution of $S_{-k}$ reflects how player $k$ conjectures that the other players will behave in the game. Since the geographic and demographic information is common knowledge known to all players, each player will form their conjectures about the other players according to their beliefs about the competitors. For simulation simplicity, we assume $S_{-k}$ has a binomial distribution with parameter 0.5. However, the theoretical analysis applies to any other distributions of $S_{-k}$.

$\mathbf{Assumption\;2}$: All service providers in the market are Bertrand competitors.

$\mathbf{Remark\;2}$: Bertrand competitors are players do not cooperate with each other. Their goal is to maximize their own utility by choosing the optimal charging station placement.

For each player, the Bayesian Nash Equilibirum (BNE) of entry actions must satisfy the best response of Equation (\ref{obj}). To solve Equation (\ref{obj}), we need to know the retail charging prices of all the service providers. Due to Bertrand competition among the service providers, the retail prices for every combination of the charging station placements are determined by the first order of conditions (FOC):
\begin{equation}\label{eq1}
\frac{\partial\Pi_1}{\partial p_1}=\sum_{n=1}^N\sum_{j=1}^{L}q_n\Phi_{j,1}^ns_{j,1}+\sum_{n=1}^N\sum_{j=1}^{L}(p_1-c_{j,1})q_n\frac{\partial\Phi_{j,1}^n}{\partial p_1}s_{j,1}=0
\end{equation}
\begin{equation}\label{eq2}
\frac{\partial\Pi_2}{\partial p_2}=\sum_{n=1}^N\sum_{j=1}^{L}q_n\Phi_{j,2}^ns_{j,2}+\sum_{n=1}^N\sum_{j=1}^{L}(p_2-c_{j,2})q_n\frac{\partial\Phi_{j,2}^n}{\partial p_2}s_{j,2}=0
\end{equation}
\begin{equation}\label{eq3}
\frac{\partial\Pi_3}{\partial p_3}=\sum_{n=1}^N\sum_{j=1}^{L}q_n\Phi_{j,3}^ns_{j,3}+\sum_{n=1}^N\sum_{j=1}^{L}(p_3-c_{j,3})q_n\frac{\partial\Phi_{j,3}^n}{\partial p_3}s_{j,3}=0
\end{equation}
where $c_{j,1},c_{j,2},$ and $c_{j,3}$ represent the LMP at each charging station candidate.

$\mathbf{Remark\;3}$: The retail prices calculated from Equations (\ref{eq1})-(\ref{eq3}) may not be the real-time prices used in practice. They are the equilibrium prices in this market under the assumption of Bertrand competition and simultaneous move game. They can be interpreted as the averaged charging prices of the service providers over a long period of time. In reality, the service providers take turns to set the retail price in respond to the prices of the competitors. It may takes a long time before the providers to reach the equilibrium prices. In addition, if the other factors changes (i.e. consumer's preference, crude oil price soaring, etc.), the previous equilibrium does not hold and new equilibrium can be calculated in a similar manner as above.

$\mathbf{Theorem\;1}$ [Strategy Decision Condition]: Under the Assumption 1 and Assumption 2, service provider $k$ will choose placement strategy $l(l=1,2,3,\cdots,2^L)$ if the type space $\Theta_k$ falls into the hypervolume specified by
\begin{equation}
\begin{aligned}
&\mathcal{H}(l)=\\
&\{\Theta_k\in \mathbb{R}_+^L:\Theta_k^{\textrm{T}}\left(S_{k,j}-S_{k,l}\right)-(\mathbb{E}R_{k,j}-\mathbb{E}R_{k,l})\\
&\;\;\;\;\;\;\;\;\;\;\;\;\;\;\;\;+w(B_{k,j}-B_{k,l})>0;\forall j \neq l\},
\end{aligned}
\end{equation}
where $S_{k,j}$ and $S_{k,l}$ denote the placement strategy $j$ and $l$, respectively. $\mathbb{E}R_{k,j}$ and $\mathbb{E}R_{k,l}$ denote the total revenue with deployment strategy $j$ and $l$, respectively.

\begin{proof}
Each service provider has $L$ candidate locations, so there are $2^L$ different placement strategies. We treat the type space as an $L$-dimensional space, and $\Theta_k=[\theta_{1,k},\theta_{2,k},\cdots,\theta_{L,k}]^{\textrm{T}}$ represents a point in this space.

By Equation (\ref{obj}), strategy $l$ is optimal if
\begin{equation}
\begin{aligned}
&\mathbb{E}[R_{k,l}]-\Theta_{k}^{\textrm{T}}S_{k,l}-wB_{k,l}>\mathbb{E}[R_{k,j}]-\\
&\;\;\;\;\;\;\;\;\;\;\Theta_{k}^{\textrm{T}}S_{k,j}-wB_{k,j};(j=1,2,\cdots,2^L, j \neq l).
\end{aligned}
\end{equation}
Rearranging the terms, we get
\begin{equation}
\begin{aligned}
&\Theta_k^{\textrm{T}}\left(S_{k,j}-S_{k,l}\right)-(\mathbb{E}R_{k,j}-\mathbb{E}R_{k,l})+\\
&\;\;\;\;\;\;\;\;\;\;\;\;\;\;w(B_{k,j}-B_{k,l})>0;(j=1,2,\cdots,2^L, j \neq l),
\end{aligned}
\end{equation}
\end{proof}

\section{Simulation Platform and Case Study}

\begin{figure}[!t]
\centerline{\includegraphics[width=3.4in]{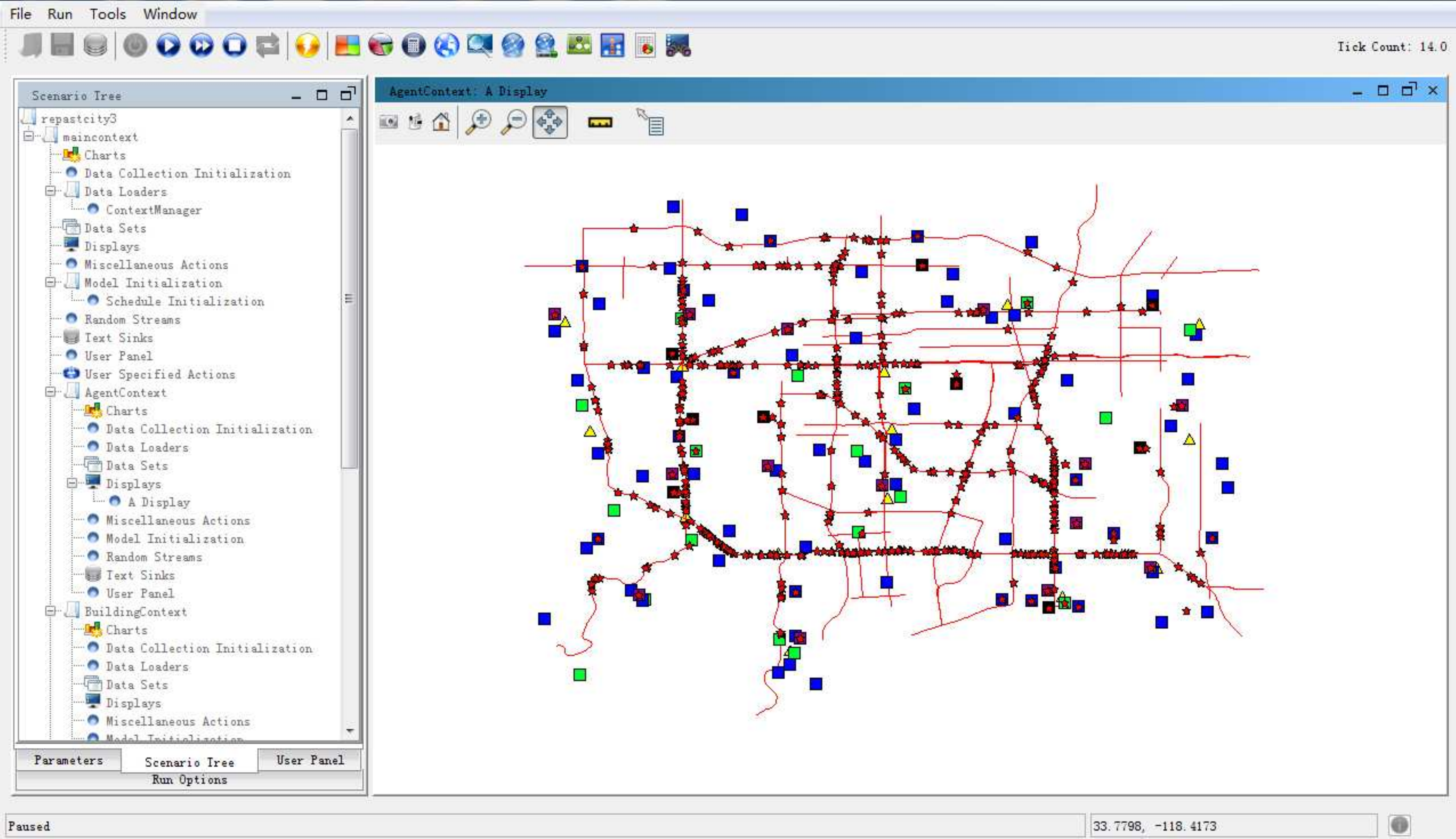}}
\center
\caption{The Screenshot of Simulation Platform}
\label{fig1}
\end{figure}

\begin{figure*}[htbp]
\centering
\subfigure[Roads and Buildings of San Pedro District]{
\label{fig4(a)} 
\includegraphics[width=3.4in]{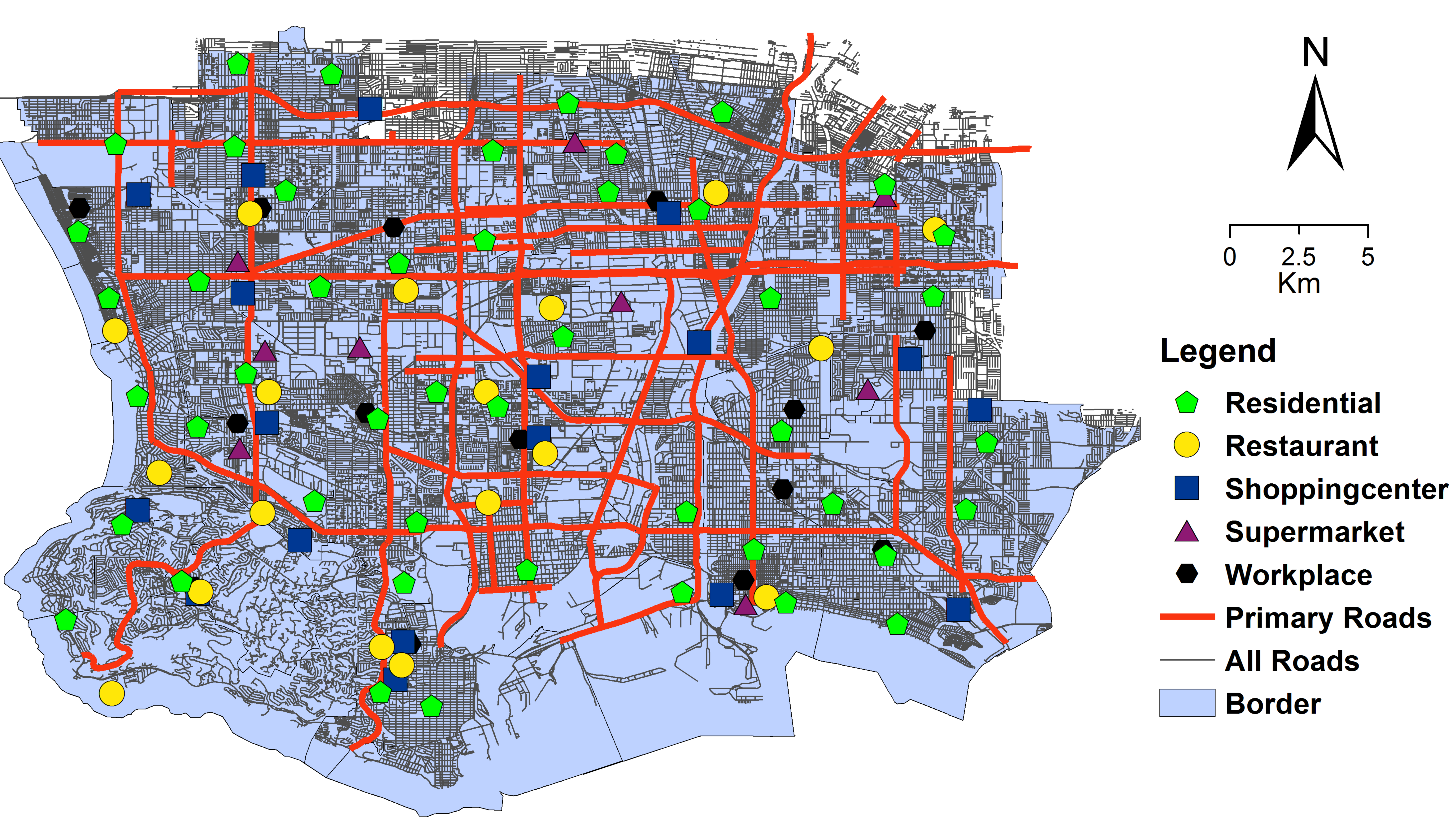}}
\subfigure[The IEEE 118-bus System]{
\label{fig4(b)} 
\includegraphics[width=3.4in]{118bus-eps-converted-to.pdf}}
\caption{Maps of San Pedro District}
\label{fig4} 
\end{figure*}

\begin{figure*}[htbp]
\centering
\subfigure[A Snapshot of EVs Movement]{
\label{fig5(a)} 
\includegraphics[width=3.4in]{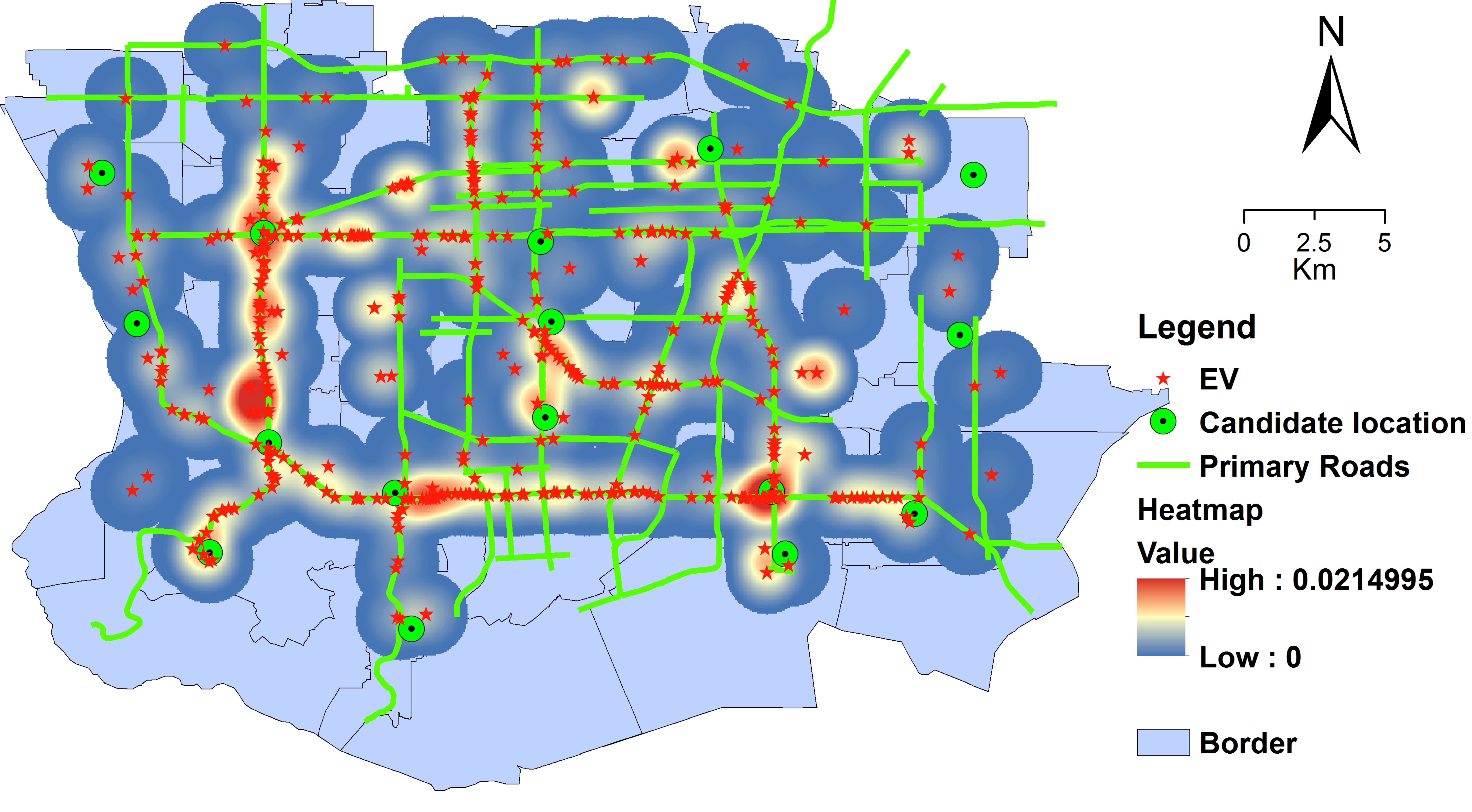}}
\subfigure[The Optimal Charging Station Locations of Three Service Providers]{
\label{fig5(b)} 
\includegraphics[width=3.4in]{result-eps-converted-to.pdf}}
\caption{Simulation Results}
\label{fig5} 
\end{figure*}

In this study, we have developed a general-purpose simulation platform (The EV Virtual City 1.0) using Repast \cite{Repast}. We conduct a case study on the San Pedro District of Los Angeles using the simulation platform. See Fig. \ref{fig1} for a screenshot of the simulation platform. See Fig.\ref{fig4(a)} for the map of San Pedro District of Los Angeles.

From the California Energy Commission website, we obtain the information of transmission line and substations in San Pedro District. This region has 107 substations in total. We use the IEEE 118-bus power system test case in our simulation with only 107 buses (substations) and 54 generators. See Fig.\ref{fig4(b)}. The base power load is calculated using the estimated residential power load in this area. For each charging station placement, we use MATPOWER to calculate the LMP of each charging station and the generating power with and without EV charging \cite{matpower}.

In the simulations, we consider three levels of EV penetration with 5000 EVs, 10000 EVs, and 15000 EVs. The corresponding EV penetration rates are 0.32\%, 0.64\%, and 0.96\%, respectively. From the 2009 National Household Travel Survey (2009 NHTS) \cite{cite14}, we calculate the distribution of travel pattern of household: (1) 27.20\% for Home-Workplace pattern, (2) 10.90\% for Home-Shopping Center pattern, (3) 18.70\% for Home-Supermarket pattern, (4) 13.10\% for Home-Restaurant (cafe, bar) pattern, (5) 30.10\% for Home-Other pattern. We model $F_{j,k}$ as i.i.d. uniform distribution in the simulations.

A snapshot of the movement of EV owners is shown in Fig. \ref{fig5(a)}. Each red star represents an EV owner (agent). The heatmap of EV owners is also plotted in this figure. The simulation results are summarized in Table \ref{ta9}. In Fig. \ref{fig5(b)}, the blue square, red square, and green square represent the Level 1 charging station, Level 2 charging station, and Level 3 charging station, respectively. The number in the square indicates that at which stage this charging station is set up.

\begin{table*}[htbp]
\center
\caption{Charging Station Placement Strategy}\label{ta9}
\begin{tabular}{cccclc}
\toprule
Stage & Level & Delay prob. & Coverage & Newly Built Stations&Total \# of Stations\\
\midrule
Stage 1&Level 1&0.281 & 2.18 & 1, 2, 3, 4, 5, 7, 8, 9, 10, 11, 12, 13, 14, 15 & 14\\
(Penetration Rate 0.32\%, & Level 2& 0.174 & 1.59 &  2, 3, 4, 7, 9, 10, 11, 14, 15 & 9 \\
5000 EVs) & Level 3&0.024&1.11 & 2, 3, 4, 7, 12, 13 & 6\\
  \bottomrule
Stage 2 & Level 1&0.292 & 3.00 & 17, 18, 20, 22, 23, 24 & 20\\
(Penetration Rate 0.64\%, & Level 2& 0.196 & 2.11 &  18, 20, 23, 24 &13 \\
10000 EVs)&Level 3&0.064&1.44 & 23, 25 & 8\\
\bottomrule
Stage 3 & Level 1&0.289 & 3.851 & 16, 19, 21, 25, 26, 27, 28, 29 & 28\\
(Penetration Rate 0.96\%,  & Level 2& 0.195 & 3.145 &  16, 19, 21, 25, 26, 28, 29, 30 & 21\\
15000 EVs)&Level 3&0.083&2.081 & 16, 19, 21, 30 & 12\\
\bottomrule
\end{tabular}
\end{table*}

From Fig. \ref{fig5} we can make four observations:
\begin{itemize}
\item The optimal charging station deployment is highly consistent with the heatmap of EV owners movement, which demonstrates that our model can adequately capture the mobility of EV owners.

\item As for the number of charging stations, Level 1 charging station is predominant over Level 2 and Level 3, probably because it takes a longer time for Level 1 to finish charging. Hence, Level 1 service provider must place more charging stations to satisfy the average delay probability constraint. On the other hand, the difference in quantity also indicates that the marketing strategies for the three service providers are different. Service provider 1 tries to place the charging stations widely across the entire area, while service provider 3 is more likely to place the charging stations at some ``hot" locations.

\item The number of charging stations does not grow linearly with the number of EVs. Their relationship seems to follow a convex curve. At stage 1 (initial stage), service providers place more charging stations. When the number of EVs doubles or triples (compared to the initial stage), service providers can just add less charging stations to satisfy the constraints. One explanation is that the high delay probability at certain busy charging stations force EV owners to choose the other idle charging stations. In other words, the charging behavior and the temporal-spatial charging demand is reshaped due to the uneven delay probability among those charging stations.

\item Service providers prefer agglomeration instead of spatial separation. The three service providers have segmented the market by offering three distinctive products (charging services) in terms of voltage, current, charging time and charging price. Through product differentiation, they significantly soften the price competition so that they do not need to spatially separate from each other to further relax competition. This observation supports the conclusion in \cite{Competition}-\cite{chargingstation} that firms do not have to maximize differentiation in every characteristic of the product. Instead, differentiation in one dominant characteristic is sufficient to alleviate price competition.
\end{itemize}

\section{Conclusions}
In this paper, we have proposed a multi-stage consumer behavior based approach to solving the problem of EV charging station placement. The nested logit model is utilized to analyze the charging behavior of EV owners. We use the Bayesian game to characterize the competition among the service providers, and by solving the game we obtain the optimal placement strategies for the service providers. In addition, we develop a simulation platform called ``The EV Virtual City 1.0" using Java, and conduct a case study of San Pedro District of Los Angeles on this platform. The simulation results show that the charging station placement is highly consistent with the traffic flow of EVs. The observations from the simulation may serve as a guideline for the local community to effectively promote and manage the EV charging market.

\end{spacing}
\end{document}